\documentclass{jfm}

\usepackage{graphicx}
\usepackage{newtxtext}
\usepackage{newtxmath}
\usepackage{natbib}
\usepackage{hyperref}
\hypersetup{
    colorlinks = true,
    urlcolor   = blue,
    citecolor  = black,
}

\newcommand{\RomanNumeralCaps}[1]
\linenumbers

\usepackage{algpseudocode}
\usepackage{algorithm}

\newcommand{\bmit}[1]{\bm{\mathit{#1}}}

\usepackage{epstopdf, epsfig}
\usepackage{bm}
\usepackage{upgreek}
\usepackage{xcolor}

\usepackage{totcount}

\newtotcounter{citnum} 
\def\oldbibitem{} \let\oldbibitem=\bibitem
\def\bibitem{\stepcounter{citnum}\oldbibitem}

\shorttitle{Modal decomposition of nonlinear interactions in wall turbulence}
\shortauthor{U. Karban et al.}

\title{Modal decomposition of nonlinear interactions in wall turbulence}

\author{U. Karban\aff{1}
  \corresp{\email{ukarban@metu.edu.tr}}, E. Martini\aff{2}, K. Oberleithner\aff{3}, A.V.G. Cavalieri\aff{4}, P. Jordan\aff{2}}

\affiliation{
\aff{1}{Department of Aerospace Engineering, Middle East Technical University, Ankara 06800, Turkey}
\aff{2}{D\'{e}partement Fluides, Thermique, Combustion, Institut Pprime, CNRS-University of Poitiers-ENSMA, France}
\aff{3}{Laboratory for Flow Instabilities and Dynamics, Technische Universit\"at Berlin, Berlin, Germany}
\aff{4}{Instituto Tecnol\'{o}gico de Aeron\'{a}utica, S\~{a}o Jos\'{e} dos Campos/SP, Brazil}
}

\begin{document}

\maketitle
\begin{abstract}
Coherent structures are found in many different turbulent flows and are known to drive self-sustaining processes in wall turbulence. Identifying the triadic interactions which generate coherent structures can provide insights beyond what is possible with linearized models.  There are infinite possible interactions that may generate a given structure. Thus, a method to systematically study those, ranking them in terms of their contribution to the structure of interest, is essential. We here use the resolvent-based extended spectral proper orthogonal decomposition (RESPOD) approach (Karban, U. \emph{et al.} 2022 Self-similar mechanisms in wall turbulence studied using resolvent analysis. \emph{Journal of Fluid Mechanics} 969, A36) to rank the triadic interactions which give rise to the dominant coherent structures in minimal Couette flows at Reynolds number 400 and 1000. Our analysis identifies that six triadic interactions dominate the most energetic coherent structure, revealing the capability of the methodology to identify and rank nonlinear interactions. The approach can be used to analyse the energy exchange in turbulent flows and may guide the construction of reduced-order models based on the interplay between different flow modes. Based on this framework, we introduce a modelling strategy where the interactions increasing or reducing the energy of a given mode are grouped as sources and sinks, respectively. The effect of the sinks is embedded in the resolvent operator by using an eddy viscosity model. The sources are used for driving this modified resolvent operator and are shown to yield accurate flow predictions at zero frequency. We discuss that this strategy can be useful when analysing nonlinear interactions or modelling forcing at high-Reynolds-number flows.


\end{abstract}

\begin{keywords}
 
\end{keywords}

\section{Introduction} \label{sec:intro}

Turbulent flows contain coherent structures that span large spatial and temporal scales. These structures are responsible for many important phenomena, ranging from sustaining the near-wall cycle \citep{hamilton_jfm_1995} in wall-bounded flows to noise generation in jets \citep{jordan_arm_2013,cavalieri_amr_2019}. It has been shown that the linear mechanisms play a major role in the generation of coherent structures \citep{ellingsen_pof_1975,landahl_jfm_1980,robinson_arm_1991,trefethen_science_1993,
hwang_jfm_2010_2,brandt_ejm_2014,schmidt_jfm_2018,pickering_jfm_2020}. A now popular approach to investigate these mechanisms is resolvent analysis, where the Navier-Stokes (N-S) equation are arranged in input-output form in the frequency domain \citep{farrell_pof_1993,jovanovic_jfm_2005,mckeon_jfm_2010,
hwang_jfm_2010,sipp_tcfd_2012,towne_jfm_2018,lesshafft_prf_2019}. Although resolvent analysis provides a dynamical framework, in most cases, it provides only a qualitative understanding of the coherent structures and associated mechanisms. It has been shown for certain flows that modelling the nonlinear fluctuations, i.e., the color of the turbulence, is essential for better prediction of these structures \citep{zare_jfm_2017,martini_jfm_2020,amaral_jfm_2021,morra_jfm_2021,
nogueira_jfm_2021,karban_jfm_2022}, particularly when developing flow models that can quantitatively predict coherent structures. 

One way to tackle the nonlinearity is to use eddy viscosity. It has been shown for many flows that adopting a frozen, i.e., time-invariant, eddy viscosity model while constructing the resolvent operator improves the prediction of coherent structures \citep{hwang_jfm_2010_2,morra_jfm_2019,
morra_jfm_2021,pickering_jfm_2021,kuhn_jfm_2021}.  {Using frozen eddy viscosity incorporates in to the resolvent operator the dissipative effects of nonlinear interactions.}

An alternative approach to model the nonlinearity is to use quasi-linear approximation \citep{malkus_jfm_1956}, where the N-S equations are split into averaged quantities and the remaining fluctuating terms. The equations for the averaged quantities are then solved directly taking into account the coupling with the fluctuation equations, while the fluctuation equations are linearised by neglecting the nonlinear fluctuating terms \citep{malkus_jfm_1956} or replacing them with a linear model \citep{farrell_jfm_2012,thomas_pof_2014,constantinou_jop_2014,
bretheim_pof_2015,
farrell_rsta_2017,bretheim_jot_2018}.

When decomposing the flow into Fourier modes, the quadratic nonlinearity of the incompressible N-S equations become triadic interactions between these modes. All these approaches involve modelling the nonlinear terms as a bulk quantity rather than tracing separately the triadic interactions of which they are formed. For a high-Reynolds-number turbulent flow, the vast number of possible interactions forming a given nonlinear term prohibits their direct modelling. There are some studies which analytically investigate triadic interactions in simple cases such as homogeneous turbulence \citep{kraichnan_jfm_1973,waleffe_pof_1992,moffatt_jfm_2014}. \cite{cheung_jfm_2014} investigated the analytical properties of triadic interactions in homogeneous, isotropic turbulence using the spectral N-S equations. They showed that the famous -5/3 decay is embedded in the N-S equations. In a recent study, \cite{cho_jfm_2018} employed the spectral turbulent kinetic energy equation to trace the energy transfer between different scales in a turbulent channel via triadic interactions. \cite{jin_prf_2021} adopted a similar approach to study the energy transfer in cylinder wake. 

Triadic interactions in turbulent flows are also investigated within the resolvent framework. The interactions between the response modes of the resolvent operator and their effect of the self-similar nature of these modes was first discussed in \cite{sharma_rsta_2017}. \cite{rosenberg_prf_2019} showed that by including the effect of triadic interactions among the dominant response modes of the resolvent operator, prediction of coherent structures can be significantly improved in oscillatory flows. This approach was followed by \cite{symon_jfm_2019} where they studied flow over airfoils, and then by \cite{symon_jfm_2021}, where they investigated the energy transfer in some minimal flow units. A formalism was provided by \cite{padovan_jfm_2020} to extend the resolvent framework to oscillatory flows, taking into account the cross-frequency interactions.  \cite{rigas_jfm_2021} used the resolvent framework together with limited triadic interactions to investigate boundary layer transition. \cite{bae_jfm_2021} investigated critical nonlinear mechanisms in Poiseuille flow, again using resolvent framework, by filtering the contribution of the dominant forcing mode to response generation.

In this study, we investigate dominant nonlinear mechanisms in wall-bounded turbulence. The complexity of all possible triadic interactions in a turbulent flow can be reduced by focusing on a certain quantity and eliminating all the non-relevant interactions. We use the resolvent-based extended spectral proper orthogonal decomposition (RESPOD) \citep{towne_aiaa_2015, karban_jfm_2022} for this purpose. RESPOD is used to rank the triadic interactions in terms of their correlation with and/or their contribution to a given observable. The method is implemented using a direct numerical simulation (DNS) of minimal Couette flow with Reynolds number 400, where the dominant coherent structures obtained by spectral proper orthogonal decomposition (SPOD) \citep{towne_jfm_2018} is chosen as the observable. In similar minimal channel configurations, \cite{bae_jfm_2021} investigated the triadic interactions contributing to the $(\alpha,\beta)=(0,2\pi/L_z)$ mode, where $\alpha$ and $\beta$ are streamwise and spanwise wavenumbers, respectively, and $L_z$ is the domain size in $z$-direction. We investigate here the triadic interactions systematically extracted using RESPOD for the same mode. By doing so, we present an approach to investigate nonlinear interactions in numerical datasets, where the effect of each triad on the observable of interest may be studied separately using the resolvent operator. This provides a quantitative analysis of the contribution of the various triads at play. The minimal Couette flow used here leads to a lower number of non-linear interactions with a few dominant coherent structures, which simplifies the analysis. The available knowledge of the dynamics of this flow allows us to demonstrate that the tool we propose does indeed identify the dominant flow interactions. 

We then combine this analysis with the use of eddy viscosity. {A static value of eddy viscosity, as discussed earlier, is used to model the overall dissipative effect of time-averaged nonlinear interactions by acting as an energy sink for the entire flow field.} This strategy is seen to be useful for improving the modelling capacity of the resolvent operator \citep{pickering_jfm_2021,morra_jfm_2021,symon_prf_2023}. However, as it already models the nonlinear interactions, which show up as the forcing term in the resolvent framework, it was unclear so far how to use eddy viscosity to make quantitative predictions when one has access to the forcing terms. 

As will be seen later in our analysis, not all the nonlinear interactions extract energy from a given flow structure. Some interactions increase the energy of the observed structure by transferring energy from other scales or being associated with amplification mechanisms such as the lift-up mechanism \citep{robinson_arm_1991,brandt_ejm_2014}. Based on this observation, we propose a modelling strategy where the nonlinear interactions are grouped as sources and sinks, i.e., the interactions that have constructive or destructive effect on the observed mode, respectively. {A similar notion of source/sink decomposition was investigated in \cite{kuhn_scitech_2022} based on turbulent kinetic energy budget.} We model the sinks using frozen eddy viscosity when constructing the resolvent operator and the sources are used to drive this modified resolvent operator. Our analysis reveals that such a decomposition of the nonlinear interactions yield quantitatively accurate representation of the flow field. We also demonstrate the relevance of this decomposition investigating the Couette flow $Re=1000$. The methods are general and may be employed in other flows of interest. With the approaches developed here we move beyond the analysis capabilities given by the resolvent operator, by analysing the non-linear terms at play in turbulence dynamics.

The remainder of the paper is structured as follows: the mathematical framework to extract triadic interactions associated with a measured quantity is explained in \S\ref{sec:math}. The details about the DNS database of the minimal Couette flow are provided in \S\ref{sec:dns}. The results of identifying the relevant triadic interactions and the energy transfer via these interactions in the minimal Couette flow are discussed in \S\ref{sec:couette}. The analysis to relate the effect of the triadic interactions with the use of eddy viscosity is presented in \S\ref{sec:eddy}. Finally, some concluding remarks are provided in \S\ref{sec:conc}.

\section{Extracting nonlinear interactions using RESPOD} \label{sec:math}
We consider the incompressible Navier-Stokes (N-S) equations in Cartesian coordinates as,
\begin{align}
\mathcal{M}\partial_t\bm{q}(\bm{x},t)=\mathcal{N}\left(\bm{q}(\bm{x},t)\right),
\end{align}
where $\bm{q}=[u\,v\,w\,p]^\top$ is the state vector, $\mathcal{N}$ denotes the nonlinear N-S operator for incompressible flows and the matrix $\mathcal{M}$ is zero for the continuity equation and identity matrix for the remaining equations. Discretisation in space and Taylor series expansion around the mean, $\overline{\bm{q}}(\bm{x})$, yields
\begin{align} \label{eq:forceintime}
\mathsfbi{M}\partial_t\bm{q}^{\prime}(\bm{x},t)-\mathsfbi{A}(\bm{x})\bm{q}^{\prime}(\bm{x},t)=\mathsfbi{B}\bm{f}(\bm{x},t),
\end{align}
where $\mathsfbi{A}(\bm{x})=\partial_q\mathcal{N}|_{\overline{\bm{q}}}$ is the linear operator obtained from the Jacobian of $\mathcal{N}$ and $\bm{f}(\bm{x},t)$ denotes all the remaining nonlinear terms, interpreted as a forcing term in the momentum equations; $\mathsfbi{M}$ denotes $\mathcal{M}$ in discrete form and $\mathsfbi{B}$ imposes zero forcing at the continuity equation. Full expressions for the operators are given in \cite{nogueira_jfm_2021}. We focus on parallel flow, i.e., a flow that is homogeneous in two directions, for instance, in $x$ and $z$, with the mean flow varying only in $y$. We can modify \eqref{eq:forceintime} to cast it in the resolvent form by applying Fourier transforms in all homogeneous dimensions and rearranging as, 
\begin{align} \label{eq:forceinfreq}
\hat{\bm{q}}(\tilde{\alpha},\tilde{\beta},{\omega})=\mathsfbi{R}(\tilde{\alpha},\tilde{\beta},{\omega})\hat{\bm{f}}(\tilde{\alpha},\tilde{\beta},{\omega}),
\end{align}
where $\tilde{\alpha}$ and $\tilde{\beta}$ are the streamwise and spanwise wavenumbers, respectively, and ${\omega}$ is the angular frequency, the hat indicates a Fourier transformed quantity and $\mathsfbi{R}(\tilde{\alpha},\tilde{\beta},{\omega})\triangleq(-i{\omega}\mathsfbi{M}-\mathsfbi{A}(\tilde{\alpha},\tilde{\beta}))^{-1}\mathsfbi{B}$ is the resolvent operator. For brevity, we drop the notation showing dependence on wavenumber, wall-normal coordinate and frequency in what follows. 

For the incompressible N-S equations, the forcing term in \eqref{eq:forceintime} is given as $\bm{f}={\bm{u}^\prime}^\top\cdot\nabla\bm{u}^\prime - \overline{{\bm{u}^\prime}^\top\cdot\nabla\bm{u}^\prime}$, where $(\cdot)$ and $()^\top$ denote dot product and transpose, respectively, and the overbar denotes averaging in time and homogeneous directions $x$ and $z$. The forcing in wavenumber-frequency space, $\hat{\bm{f}}_{\bm{k}}$, is then obtained via a convolution,
\begin{align} \label{eq:conv}
\hat{\bm{f}}_{\bm{k}}=\sum_{{\bm{i}}}\hat{\bm{u}}_{\bm{i}}^\top\cdot\nabla\hat{\bm{u}}_{\bm{k-i}},
\end{align}
where ${\bm{i}}=(\tilde{\alpha}_i,\tilde{\beta}_i,\omega_i)$, and ${\bm{k}}=(\tilde{\alpha}_k,\tilde{\beta}_k,\omega_k)$ denote wavenumber-frequency combinations, and summation over ${\bm{i}}$ implies a nested summation over $\tilde{\alpha}_i$, $\tilde{\beta}_i$ and $\omega_i$. Here, we consider that $\omega$ is discretised. Note that \eqref{eq:conv} is valid assuming that the triplet ${\bm{k}}$ contains at least one non-zero element, such that the averaged term in $\bm{f}$ has no contribution.

The RESPOD method, adapted from extended proper orthogonal decomposition \citep{boree_eif_2003,hoarau_prf_2006}, finds, for a given observable, all structures in a `target' event that are correlated to the SPOD modes of the observable. Here we choose the target event to be the nonlinear interactions, which give rise to the forcing terms in the resolvent framework, as in \cite{towne_aiaa_2015} and \cite{karban_jfm_2022}. The goal is to map the triadic interactions underpinning the dominant coherent structures of the flow.

The SPOD modes of the flow field can be estimated using the ensemble matrix of realisations, through the eigendecomposition,
\begin{align}
\hat{\mathsfbi{Q}}_{\bm{k}}^H\mathsfbi{W}\hat{\mathsfbi{Q}}_{\bm{k}}=\hat{\bmit{\Theta}}_{\bm{k}}{\bmit{\Lambda}}_{\bm{k}}\hat{\bmit{\Theta}}_{\bm{k}}^H,
\end{align}
and the SPOD modes are obtained from $\hat{\bmit{\Theta}}_{\bm{k}}$ as,
\begin{align} \label{eq:resmain}
{\bmit{\Psi}}_{\bm{k}}=\hat{\mathsfbi{Q}}_{\bm{k}}\hat{\bmit{\Theta}}_{\bm{k}}{\bmit{\Lambda}}^{-1/2}_{\bm{k}},
\end{align}
where $\hat{\mathsfbi{Q}}_{\bm{k}}\triangleq[{{}\hat{\bm{q}}_{\bm{k}}}_{(1)} \, {{}\hat{\bm{q}}_{\bm{k}}}_{(2)} \cdots {{}\hat{\bm{q}}_{\bm{k}}}_{(P)}]$ denotes the ensemble matrix for different realisations of $\hat{\bm{q}}_{\bm{k}}$ with $P$ being the total number of realisations, ${\bmit{\Psi}}_{\bm{k}}$ and ${\bmit{\Lambda}}_{\bm{k}}$ are SPOD modes and their associated eigenvalues, respectively (see \cite{towne_jfm_2018}), and $\mathsfbi{W}$ is a positive-definite matrix of quadrature gains along $y$, which is discretised. The SPOD modes in the columns of ${\bmit{\Psi}}_{\bm{k}}$ are the optimal orthonormal basis for the realisations of the observable ${\hat{\bm{y}}}_{\bm{k}}$.

In \cite{karban_jfm_2022}, it was shown that the coefficient matrix $\hat{\bmit{\Theta}}_{\bm{k}}$ can be used to extract the part in the forcing that is correlated with the observed SPOD mode as
\begin{align} \label{eq:respoddef}
{\mathsfbi{X}}_{\bm{k}} = \hat{\mathsfbi{F}}_{\bm{k}}\hat{\bmit{\Theta}}_{\bm{k}}{\bmit{\Lambda}}^{-1/2}_{\bm{k}},
\end{align}
where, $\hat{\mathsfbi{F}}_{\bm{k}}$ is the ensemble matrix of $\hat{\bm{f}}_{\bm{k}}$. The RESPOD forcing mode $\mathsfbi{X}_{\bm{k}}$ satisfies
\begin{align} \label{eq:respod}
{\bmit{\Psi}}_{\bm{k}}={\mathsfbi{R}}_{\bm{k}}{\mathsfbi{X}}_{\bm{k}},
\end{align}
i.e., the RESPOD forcing mode excites precisely the SPOD mode via the resolvent operator. As discussed in \cite{karban_jfm_2022} and \cite{karban_jfm_2023}, the RESPOD mode includes the part of the forcing that is correlated to the SPOD mode of interest. 

Substituting into \eqref{eq:respoddef} the expansion in \eqref{eq:conv}, which shows the triadic interactions summing up to yield the forcing $\hat{\mathsfbi{F}}_{\bm{k}}$, one can compute the triadic interactions correlated with the observable as
\begin{align} \label{eq:espodnonlin}
\hat{\mathsfbi{F}}_{\bm{k}}\hat{\bmit{\Theta}}_{\bm{k}}{\bmit{\Lambda}}^{-1/2}_{\bm{k}}=\sum_{{\bm{i}}}\left(\hat{\mathsfbi{U}}_{\bm{i}}^\top\cdot\nabla\hat{\mathsfbi{U}}_{{\bm{k-i}}}\right)^\top\hat{\bmit{\Theta}}_{\bm{k}}{\bmit{\Lambda}}^{-1/2}_{\bm{k}},
\end{align}
where $\hat{\mathsfbi{U}}$ denotes the ensemble matrix of $\hat{\bm{u}}$. Defining
\begin{align} \label{eq:espodnonlin2}
\mathsfbi{X}_{{\bm{i}},{\bm{k-i}}}\triangleq\left(\hat{\mathsfbi{U}}_{\bm{i}}^\top\cdot\nabla\hat{\mathsfbi{U}}_{{\bm{k-i}}}\right)^\top\hat{\bmit{\Theta}}_{\bm{k}}{\bmit{\Lambda}}^{-1/2}_{\bm{k}},
\end{align}
the correlated forcing $\mathsfbi{X}_{\bm{k}}$ can be decomposed as,
\begin{align} \label{eq:espodnonlin3}
\mathsfbi{X}_{\bm{k}}=\sum_{{\bm{i}}}\mathsfbi{X}_{\bm{i},\bm{k-i}}=\sum_{{\bm{i}}}\left(\hat{\mathsfbi{U}}_{\bm{i}}^\top\cdot\nabla\hat{\mathsfbi{U}}_{{\bm{k-i}}}\right)^\top\hat{\bmit{\Theta}}_{\bm{k}}{\bmit{\Lambda}}^{-1/2}_{\bm{k}}.
\end{align}
{In this study, we focus on the optimal SPOD mode, i.e., the column in $\bmit{\Psi}_{\bm{k}}$ corresponding the largest eigenvalue, and the associated forcing mode, defined as $\bm{\psi}_{\bm{k}}$ and $\bm{\chi}_{\bm{k}}$, respectively. Following the notation in \eqref{eq:espodnonlin3}, individual triadic interactions, which yield $\bm{\chi}_{\bm{k}}$ when summed over $\bm{i}$ are denoted with $\bm{\chi}_{\bm{i},\bm{k-i}}$. Similarly, the response to a single triadic interaction is denoted as $\bm{\psi}_{\bm{i},\bm{k-i}}=\mathsfbi{R}_{\bm{k}}\bm{\chi}_{\bm{i},\bm{k-i}}$, which also implies that summation of $\bm{\psi}_{\bm{i},\bm{k-i}}$ over $\bm{i}$ results in $\bm{\psi}_{\bm{k}}$.} 

We define the energy as,
\begin{align} \label{eq:wnorm}
\|(\cdot)\|^2=\varepsilon\{(\cdot)^H\mathsfbi{W}(\cdot)\},
\end{align}
where the superscript $H$ indicates Hermitian transpose, and $\varepsilon\{\cdot\}$ denotes the expectation operator. In what follows, $\varepsilon\{\cdot\}$ corresponds to time-averaging for time-dependent structures, and to ensemble averaging for Fourier realisations in the frequency space. The energy of $\bm{\chi}_{{\bm{i}},{\bm{k-i}}}$, denoted by $\|{\bm{\chi}}_{{\bm{i}},{\bm{k-i}}}\|^2$, for all ${\bm{i}}$ shows the correlation map of the nonlinear interactions related to the observed SPOD mode, $\bm{\psi}_{\bm{k}}$. One can instead investigate $\|\bm{\psi}_{{\bm{i}},{\bm{k-i}}}\|^2$, which provides the contribution of a triadic interaction to the optimal SPOD mode of the observed state, as suggested by \eqref{eq:respod} and \eqref{eq:espodnonlin3}. By removing or including terms in the sum in equation \eqref{eq:espodnonlin3}, one is able to inspect the contributions of each triad $\bm{i}$ to the observable.

\section{Database of the minimal Couette flow} \label{sec:dns}
The use of RESPOD for detection of `important' nonlinear interactions associated with a specific measurement is tested on a minimal Couette flow \citep{hamilton_jfm_1995}, similar to that investigated by \cite{nogueira_jfm_2021}. The simulations are performed using the ‘ChannelFlow’ code, a pseudo-spectral incompressible flow solver using a Fourier-Chebyshev discretisation in the wall-parallel and wall-normal directions, respectively  (see www.channelflow.ch for details). The dimensions of the minimal box are $(L_x,L_y,L_z)=(1.75\pi h,2h,1.2\pi h)$, where the subscripts $x$, $y$ and $z$ denote the streamwise, wall-normal and spanwise directions, and $h$ is the channel half-height. These are the minimal dimensions to sustain turbulence in Couette flow at low Reynolds number, as studied by \cite{hamilton_jfm_1995}. For simpler notation, wavenumbers will be presented in integers defined as $\alpha=\tilde{\alpha}L_x/2\pi$ and $\beta=\tilde{\beta}L_z/2\pi$. 

The domain was discretised as $(n_x,n_y,n_z)=(32,65,32)$ with a dealiasing factor of 3/2 in the wall-parallel directions. The channel walls move with wall velocity, $\pm U_w$ yielding a Reynolds number, $Re=400$ based on $U_w$ and $h$, corresponding to a friction Reynolds number, $Re_\tau\approx34$. Once the initial transients disappeared, the flow data was stored for 7000 convective units with a sampling rate, $\Delta t=0.25$. Temporal data is transformed into frequency space using blocks of 2048 time steps with 50\% overlapping and using a second-order exponential windowing function given in \cite{martini_arxiv_2019}. While computing the forcing data, the correction due to using windowing functions is implemented as described in \cite{martini_arxiv_2019} and \cite{nogueira_jfm_2021}. We verified that the forcing acting on the resolvent operator accurately yields the response, however, the comparison is not shown here for brevity.

\begin{figure}
  \centerline{\resizebox{\textwidth}{!}{\includegraphics{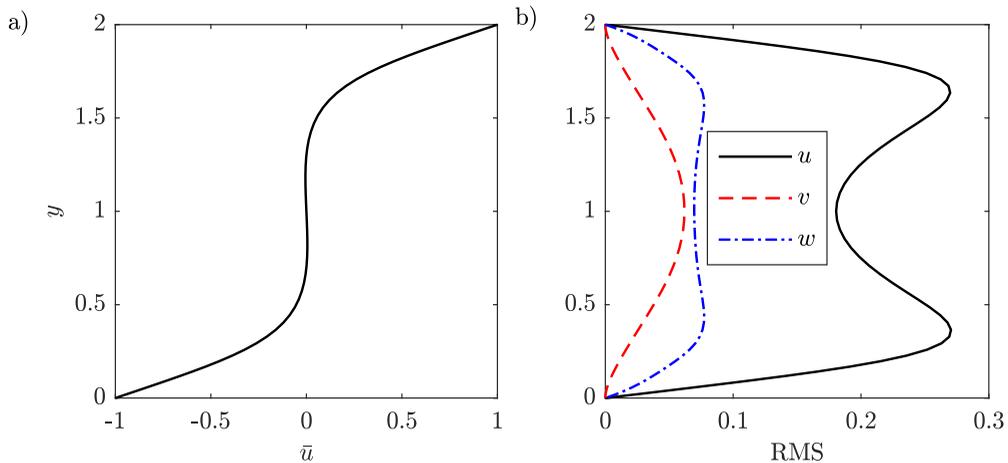}}}
  \caption{Mean (a) and the RMS (b) profiles of the velocity components, $u$ (black solid), $v$ (red dashed) and $w$ (blue dash-dotted) along the wall-normal direction.}
\label{fig:meanrms}
\end{figure}

Figure \ref{fig:meanrms} presents the profiles for the mean and the root-mean-square (RMS) of the velocity components, $u$, $v$ and $w$ in the streamwise, wall-normal and spanwise directions, respectively, along the wall-normal direction, $y$. We see that the mean flow deviates from the laminar solution given by $(y-1)$ due to nonlinear interactions between turbulent fluctuations. The RMS plots indicate that the fluctuations in $u$ peak around $y=1.5$ and $y=0.5$. A similar but smaller double-peak structure is seen in the RMS of $w$ with the peaks occurring at the same wall-normal positions. The RMS of $v$ peaks around the centre at an amplitude slightly lower than that of $w$.

\begin{figure}
  \centerline{\resizebox{\textwidth}{!}{\includegraphics{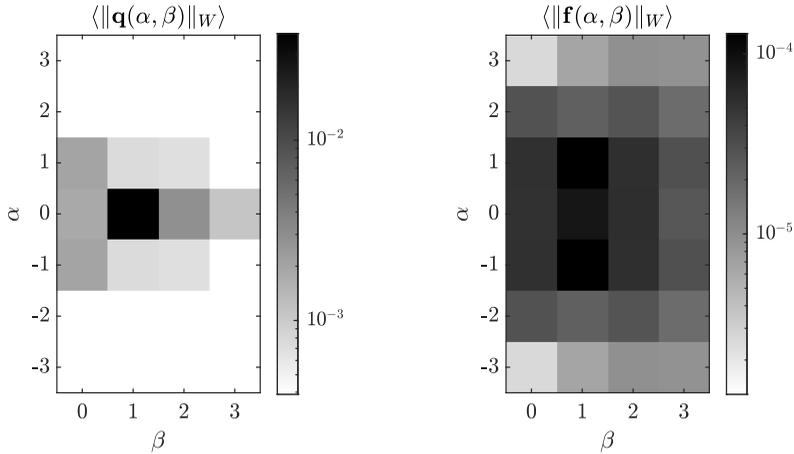}}}
  \caption {Time average of the energy of flow structures (a) and the associated forcing (b) at different wavenumber pairs. The color-scale ranges two orders of magnitude for both plots.}
\label{fig:resprms}
\end{figure} 

Minimal Couette flow is known to be dominated most of the time by rolls and streaks spanning the entire computational domain, corresponding to $(\alpha,\beta)=(0,1)$. Occasionally wavy disturbances with $\alpha=1$ appear after streak instability and breakdown, and subsequently non-linear interactions among such ``waves'' lead to the formation of new rolls, restarting the process \citep{hamilton_jfm_1995,hall_jfm_2010}. Figure \ref{fig:resprms} shows the time-averaged energy contained in each wavenumber pair and the associated forcing. We see that the mode pair $(\alpha,\beta)=(0,1)$, related to streaks and rolls, contains most of the fluctuation energy ($\sim$75\%). In contrast, the modes $(\pm1,0)$, related to waves, and $(0,2)$, which we will refer to as roll-streak harmonic, contain slightly less than 5\% of the total energy, and all the other mode pairs have less than 2\%. The energy among the forcing modes is more evenly distributed. The dominant modes are $(\pm1,1)$, each containing $\sim$13\% of the total forcing energy. 

\begin{figure}
  \centerline{\resizebox{\textwidth}{!}{\includegraphics{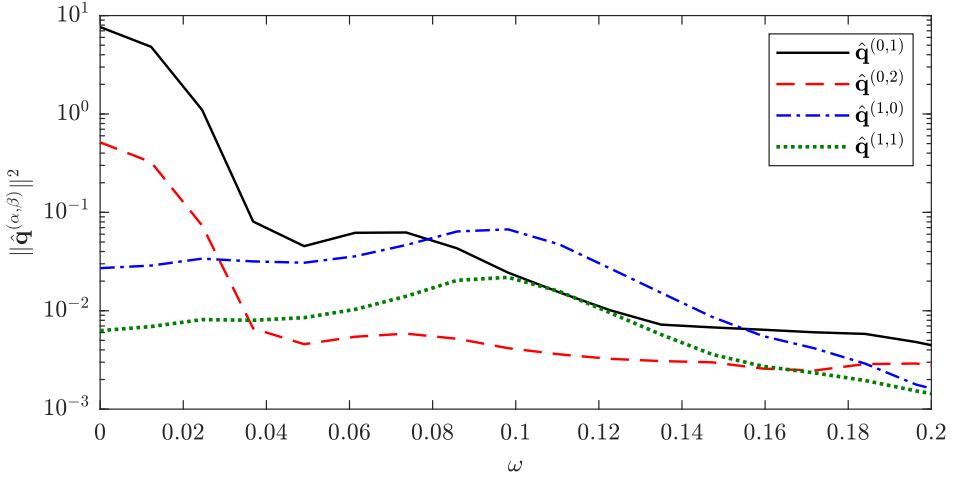}}}
  \vspace{-20pt}
  \caption{PSDs of $\hat{\bm{q}}^{(0,1)}$ (blue), $\hat{\bm{q}}^{(0,2)}$ (orange), $\hat{\bm{q}}^{(1,0)}$ (yellow) and $\hat{\bm{q}}^{(1,1)}$ (violet) integrated over the wall-normal direction.}
\label{fig:resppsd}
\end{figure}

\begin{figure}
  \centerline{\resizebox{\textwidth}{!}{\includegraphics{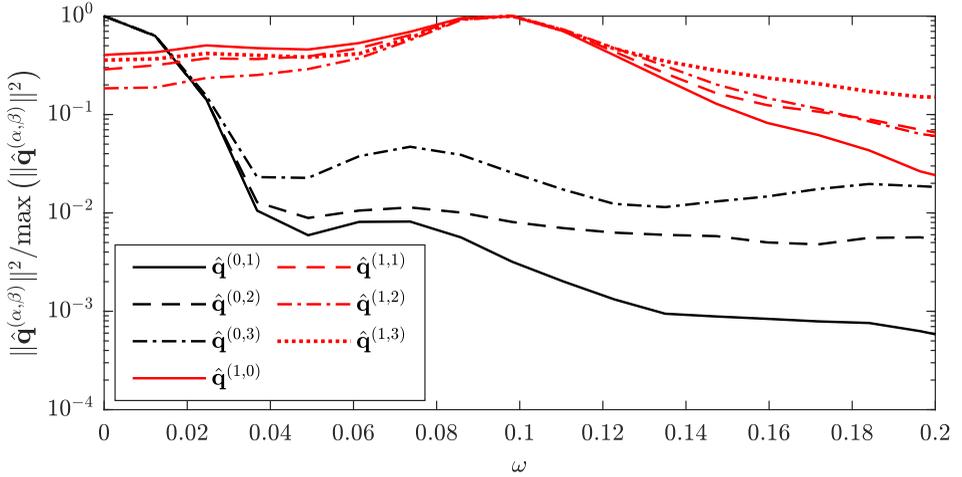}}}
  \vspace{-20pt}
  \caption{PSDs of $\hat{\bm{q}}^{(0,\{1,3\})}$ (black; solid, dashed, dash-dotted, respectively), and $\hat{\bm{q}}^{(1,\{0,3\})}$ (red; solid, dashed, dash-dotted, dotted, respectively) integrated over the wall-normal direction and normalised with respect to the peak value of each mode.}
\label{fig:resppsdnrm}
\end{figure}

Figure \ref{fig:resppsd} shows the power spectral density (PSD), integrated along the wall-normal direction, of the velocity field $\bm{q}$ at wavenumber pairs $(\alpha,\beta)=(0,1)$, $(0,2)$, $(1,0)$ and $(1,1)$. Although the oblique-wave mode $(\alpha,\beta)=(1,1)$ is energy-wise insignificant, it plays a critical role for transfer of energy to $(\alpha,\beta)=(0,1)$ mode, as will be shown later, and hence is included here. We see that the streamwise-constant modes peak around the zero frequency, which is expected due to their quasi-steady nature, while the wave modes $(1,0)$ and $(1,1)$ have their peak around ${\omega}\approx0.1$, leading to a phase speed of $c^+\triangleq{\omega}^+/\tilde{\alpha}^+=\pm1$ in wall units (negative values arise if frequency or wavenumber is negative) corresponding to $\sim10\%$ of wall velocity. The shape of the spectra is observed to be similar for the modes that have the same streamwise wavenumber. This trend can be more clearly seen in figure \ref{fig:resppsdnrm}, where the integrated PSDs normalised with respect to the peak value are plotted for different modes. We see two different families of PSD distributions for the two streamwise wavenumbers, $\alpha=0$ and $\alpha=1$, respectively.

\begin{figure}
  \centerline{\resizebox{\textwidth}{!}{\includegraphics{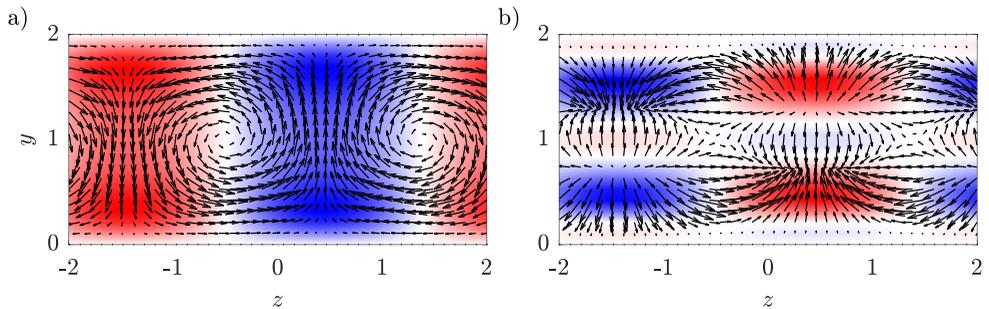}}}
  \vspace{-20pt}
  \caption{The optimal SPOD mode $\bm{\psi}_{\bm{k}}$ (a) and the associated forcing $\bm{\chi}_{\bm{k}}$ (b) reconstructed in the $y$-$z$ plane for the mode $\bm{k}=(\alpha,\beta,\omega)=(0,1,0)$. The color plot indicates the streamwise component and the arrows show the spanwise and wall-normal components.}
\label{fig:frcrespmode}
\end{figure} 

\begin{figure}
  \centerline{\resizebox{\textwidth}{!}{\includegraphics{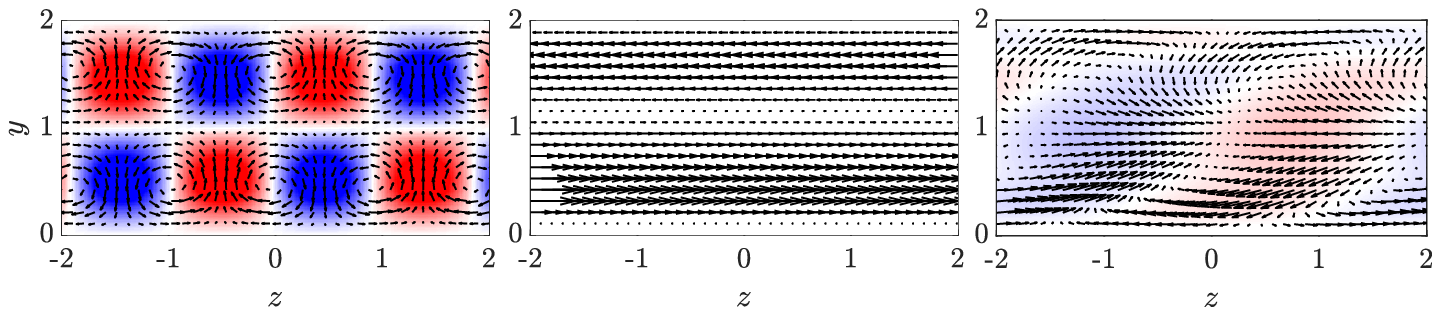}}}
  \vspace{-5pt}
  \caption{The optimal SPOD modes reconstructed in the $y$-$z$ plane for the modes $(\alpha,\beta,\omega)=(0,2,0)$ (left), $(1,0,0.1)$ (center) and $(1,1,0.1)$ (right). The color plot indicates the streamwise component and the arrows show the spanwise and wall-normal components.}
\label{fig:3respmodes}
\end{figure} 

We now focus on the most energetic mode $(\alpha,\beta)=(0,1)$ at its peak-energy frequency, $\omega=0$. The optimal response mode and the associated forcing, $\bm{\psi}_{\bm{k}}$ and ${\bm{\chi}}_{\bm{k}}$, respectively, are shown in figure \ref{fig:frcrespmode}. The optimal response consists of streaks and rolls. Given that the upper and lower walls have positive and negative mean velocities, respectively, the phase relation between streaks and rolls is reminiscent of the lift-up mechanism \citep{brandt_ejm_2014}. This is further supported regarding the associated forcing mode. At the spanwise positions where the streamwise vortices are located, the forcing is mainly located near the walls aligned with the $z$-direction and with the $y$-direction to the left and right of the vortices, causing a moment to generate the streamwise vortices. These vortices then generate streaks by carrying the high- and low-velocity structures near the upper and lower walls, respectively, towards the channel centre. Note that the forcing component in the streamwise direction is in opposite phase to the streaks seen in the response. This indicates that the streaks are generated by the lift-up mechanism despite the counteracting effect of the streamwise forcing, as previously reported by \cite{nogueira_jfm_2021}. The response generation at this triplet can therefore be considered suboptimal.

{We also plot the optimal response fields for the modes $(\alpha,\beta,\omega)=(0,2,0)$, $(1,0,0.1)$ and $(1,1,0.1)$, respectively, in figure \ref{fig:3respmodes}. Each mode is shown at its peak frequency (see figure \ref{fig:resppsd}). The response field contains streaks and rolls for the mode $(0,2,0)$ as in the roll-streak mode $(0,1,0)$, but with doubled periodicity, and thus, is called roll-streak harmonic. The mode $(1,0,0.1)$ is dominated by its spanwise component, leading to a wave mode. Finally, the response field for the mode $(1,1,0.1)$ contains some oblique wave structures tilted with the mean flow.}

\section{Nonlinear interactions in the minimal Couette flow} \label{sec:couette}
\subsection{Extracting important triadic interactions} \label{subsec:tri}
\begin{figure}
  \centerline{\resizebox{\textwidth}{!}{\includegraphics{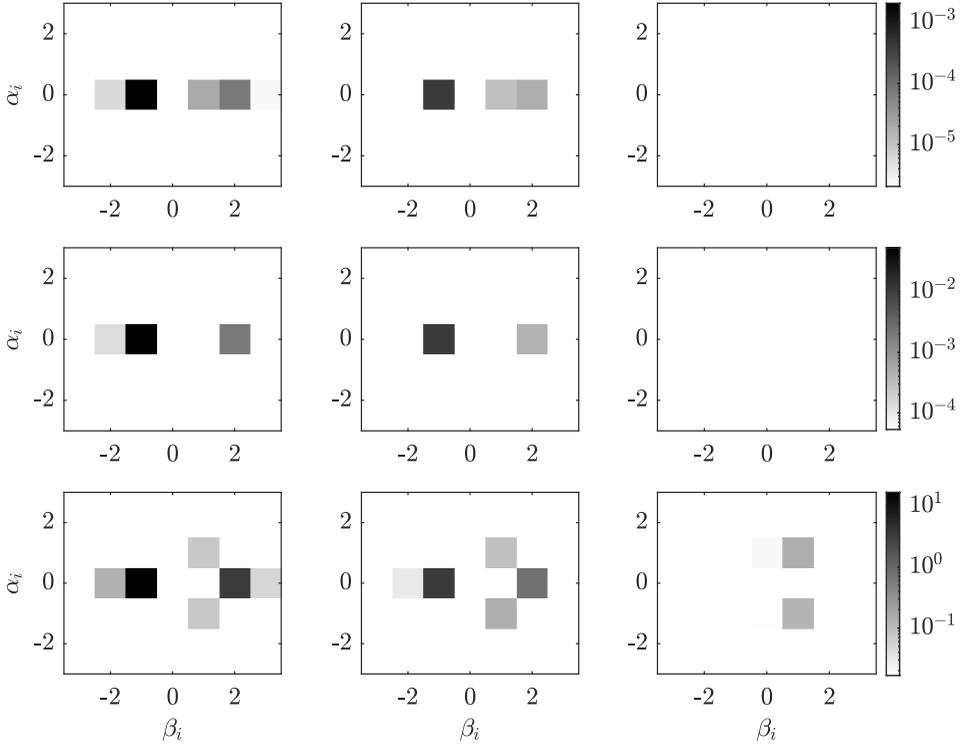}}}
  \vspace{-5pt}
  \caption{Amplitude maps of $\|\bm{u}_{\bm{i}}^\top\cdot\nabla\bm{u}_{{\bm{k-i}}}\|^2$ (top), $\|\bm{\chi}_{{\bm{i}},{\bm{k-i}}}\|^2$ (middle), and $\|\bm{\psi}_{{\bm{i}},{\bm{k-i}}}\|^2$ (bottom) obtained at $\omega_i=0$ (left), $\omega_i=0.05$ (center) and $\omega_i=0.1$ (right), for the mode ${\bm{k}}=(\alpha_k,\beta_k,\omega_k)=(0,1,0)$. Only the modes ${\bm{i}}$ are shown while the complementary modes ${\bm{k-i}}$ are selected to yield ${\bm{k}}=(0,1,0)$.}
\label{fig:espmap3freq}
\end{figure}

The maps showing the energy of the nonlinear interactions contributing to the dominant mode, ${\bm{k}}=(\alpha_k,\beta_k,\omega_k)=(0,1,0)$ are given in figure \ref{fig:espmap3freq}. Different columns compares the maps $\|\hat{\bm{u}}_{\bm{i}}\nabla\hat{\bm{u}}_{\bm{k-i}}\|^2$, $\|\bm{\chi}_{{\bm{i}},{\bm{k-i}}}\|^2$ and $\|\bm{\psi}_{{\bm{i}},{\bm{k-i}}}\|^2$, which correspond respectively to energies of the direct triadic interactions, the interactions correlated with the optimal SPOD mode, and the response to the latter obtained by driving the resolvent operator. Note that only the triplet ${\bm{i}}=(\alpha_i,\beta_i,\omega_i)$ is shown, where for each ${\bm{i}}$, there exists a ${\bm{k-i}}$ such that the nonlinear interaction between $\bm{i}$ and $\bm{k-i}$ yields ${\bm{k}}=(0,1,0)$. Starting with $\omega_i=0$, we see that the interaction between the roll-streak mode, $\bm{i}=(0,-1,0)$ and its complementary, roll-streak harmonic $\bm{k-i}=(0,2,0)$, is dominant in all three maps, indicating that the interaction is large in amplitude, highly correlated to the dominant mode, and generates the response with the largest amplitude. We also observe large amplitude for the interaction $(0,2,0)+(0,-1,0)$, which involves the same structures with the previous one, but with the gradient operator acting on the roll-streak mode $(0,-1,0)$. The interactions involving wave modes $(\pm1,1,0)+(\mp1,0,0)$, although not yielding a large forcing component (low amplitudes at the first two rows of figure \ref{fig:espmap3freq}), are seen to be present in the response map (third row of figure \ref{fig:espmap3freq}), implying that these modes efficiently drive the observable. For non-zero frequencies $\omega_i = 0.05$ and 0.1, we observe that the contribution of streamwise-constant modes with $\alpha_i = 0$ decreases with increasing $\omega$, whereas wave modes with $\alpha_i= 1$ drive an increasingly stronger response for higher frequencies, which may be attributed to the different frequency content of streamwise-constant and wavy modes, explored in figures \ref{fig:resppsd} and \ref{fig:resppsdnrm}.

\begin{figure}
  \centerline{\resizebox{\textwidth}{!}{\includegraphics{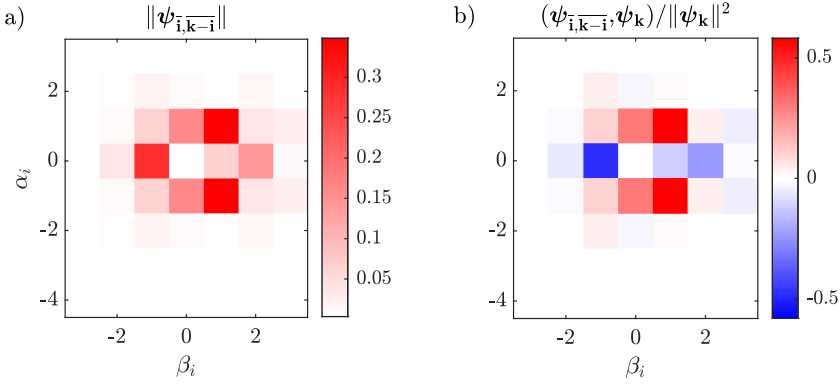}}}
  \caption{a) Energy map of the response generated by the triadic interactions at all frequencies added together. b) The map of normalised inner product between the optimal SPOD mode and the response generated by $\bm{\psi}_{\overline{\bm{i}},\overline{\bm{k-i}}}$. Both maps are shown for the roll-streak mode, $(\alpha_k,\beta_k,\omega_k)=(0,1,0)$.}
\label{fig:espmapallfreq}
\end{figure} 
{To investigate the overall contribution to the dominant mode $(0,1,0)$ via a given wavenumber pair denoted by $\overline{\bm{i}}=(\alpha_i,\beta_i)$ and its complementary $\overline{\bm{k-i}}$, we define the forcing mode summed over the frequency index, $\omega_i$, as
\begin{align}
	\bm{\chi}_{\overline{\bm{i}},\overline{\bm{k-i}}}=\sum_{\omega_i} \bm{\chi}_{{\bm{i}},{\bm{k-i}}},
\end{align} 
which amounts to adding the nonlinear interactions between all different frequency combinations. The associated response is defined as  $\bm{\psi}_{\overline{\bm{i}},\overline{\bm{k-i}}}\triangleq\tilde{\mathsfbi{R}}_{\bm{k}}\bm{\chi}_{\overline{\bm{i}},\overline{\bm{k-i}}}$.} Similar to the energy maps shown in figure \ref{fig:espmap3freq}, the map of $\|\bm{\psi}_{\overline{\bm{i}},\overline{\bm{k-i}}}\|$ is plotted in figure \ref{fig:espmapallfreq}-a, which shows that the response generation is dominated by six interactions: two streamwise-constant, which are $(0,\{-1,2\})+(0,\{2,-1\})$ involving the roll-streak and roll-streak harmonic modes, and four streamwise-periodic over $L_x$, which are $(\pm1,\{0,1\})+(\mp1,\{1,0\})$ involving wave modes. Note that here and in what follows, we use curly brackets for short hand notation of multiple modes. For instance, $(0,\{-1,2\})$ denotes the modes $(0,-1)$ and $(0,2)$.

Besides the magnitude of the response to a given triadic interaction, it is important to evaluate how it contributes to the overall response. As shown in \cite{nogueira_jfm_2021} and \cite{morra_jfm_2021}, different forcing components can interfere destructively. In what follows we propose a measure to identify which interactions are constructive, amplifying a given mode, or destructive, saturating or damping it. As a measure of constructiveness/destructiveness of a given interaction, we calculate the inner product between the response to a triadic interaction, $\bm{\psi}_{\overline{\bm{i}},\overline{\bm{k-i}}}$, and the optimal SPOD mode, $\bm{\psi}_{\bm{k}}$, 
\begin{align} \label{eq:qinner}
\langle\bm{\psi}_{\overline{\bm{i}},\overline{\bm{k-i}}},\bm{\psi}_{\bm{k}}\rangle\triangleq \varepsilon\left\lbrace\Re(\bm{\psi}_{\overline{\bm{i}},\overline{\bm{k-i}}}^H \mathsfbi{W}\bm{\psi}_{\bm{k}})\right\rbrace,
\end{align}
where $\Re(\cdot)$ denotes taking the real part. An \emph{interaction map} is obtained by calculating \eqref{eq:qinner} for each wavenumber pair and normalising the result with $\|{{}\hat{\bm{\psi}}_{\bm{k}}}\|^2$, which shows a normalised projection, and thus the constructive/destructive role of each triadic interaction in generating the response. The resulting map is shown in figure \ref{fig:espmapallfreq}-b. Note that the interaction map should sum up to 1, i.e., the sum of all destructive and constructive interactions lead to the mode observed in the system. {This can be shown via the following identity;
	\begin{align}
		\sum_{\overline{\bm{i}}}\langle\bm{\psi}_{\overline{\bm{i}},\overline{\bm{k-i}}},\bm{\psi}_{\bm{k}}\rangle = \left\langle\sum_{\overline{\bm{i}}}\bm{\psi}_{\overline{\bm{i}},\overline{\bm{k-i}}},\bm{\psi}_{\bm{k}}\right\rangle = \langle\bm{\psi}_{\bm{k}},\bm{\psi}_{\bm{k}}\rangle=1.
	\end{align}
	The sum of individual contributions being 1 implies that only the real part of the inner product shown in \eqref{eq:qinner} can contribute to the mode $\bm{\psi}_{\bm{k}}$ on average, thus, a real operator is used in the definition of the inner product.} 

The analysis reveals that the contributions from the interactions $(0,\{-1,2\})+(0,\{2,-1\})$ decrease the response energy, implying a destructive interference between these interactions and the remaining ones. The interactions involving wave modes $(\pm1,\{0,1\})+(\mp1,\{1,0\})$ , on the other hand, cause the response energy to increase, implying a constructive effect. 

\begin{figure}
  \centerline{\resizebox{\textwidth}{!}{\includegraphics{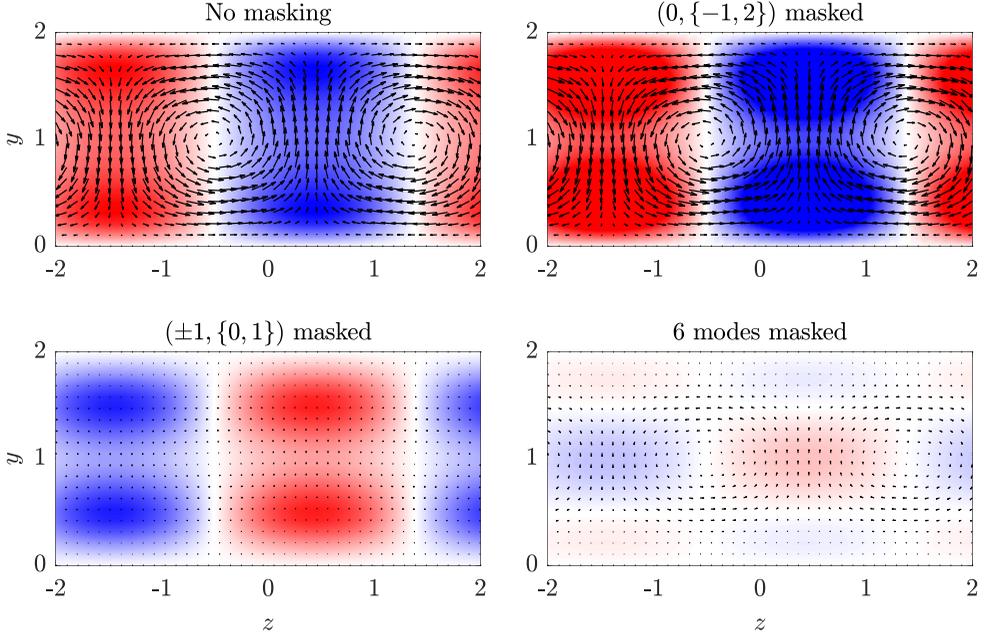}}}
  \caption{Velocity field corresponding to the optimal SPOD mode at $(\alpha_k,\beta_k,\omega_k)=(0,1,0)$ in the $y$-$z$ plane. Top-left: the entire response; top-right: the response obtained by masking the interactions between the modes $(\alpha_i,\beta_i)=(0,\{-1,2\})$ and their complementary modes; bottom-left: the response obtained by masking the interactions between the modes $(\alpha_i,\beta_i)=(\pm1,\{0,1\})$ and their complementary modes; bottom-right: the response obtained by masking the interactions between the modes $(\alpha_i,\beta_i)=(0,\{-1,2\})$ and their complementary modes as well as the interactions between the modes $(\alpha_i,\beta_i)=(\pm1,\{0,1\})$ and their complementary modes.}
\label{fig:velrecmask}
\end{figure} 

The effect of these interactions on the response field is shown in figure \ref{fig:velrecmask} by masking these interactions, i.e., subtracting the contributions from the designated interactions from the overall response computed by the resolvent. We see that masking the interactions $(0,\{-1,2\})+(0,\{2,-1\})$ mainly affects the streaks causing an increase in their amplitude, while the roll remains nearly unchanged. Masking the interactions $(\pm1,\{0,1\})+(\mp1,\{1,0\})$ almost completely eliminates the streamwise vortices, which also causes the lift-up effect to be eliminated. This results in streaks with smaller amplitude and reversed phase. This result is consistent with models of self-sustaining process in wall turbulence, where rolls are excited by non-linear interactions involving waves with non-zero $\alpha$ \citep{hamilton_jfm_1995,hall_jfm_2010}. Remember that in the RESPOD forcing mode shown in figure \ref{fig:frcrespmode}, the streamwise component counteracts the lift-up mechanism forced by the spanwise components. These results, when combined, imply that the streamwise and spanwise components in the RESPOD forcing mode, $\bm{\chi}_{\bm{k}}$, are mainly constructed by the nonlinear interaction groups $(0,\{-1,2\})+(0,\{2,-1\})$ and $(\pm1,\{0,1\})+(\mp1,\{1,0\})$, respectively. Masking $(\pm1,\{0,1\})+(\mp1,\{1,0\})$ causes the lift-up mechanism, which is an efficient means to generate streaks via streamwise vortices, to disappear. The remaining streamwise component in $\bm{\chi}_{\bm{k}}$ is mainly constructed by $(0,\{-1,2\})+(0,\{2,-1\})$ and generates streaks with negative phase, reducing the amplitude of the streaks generated by the lift-up mechanism. This elucidates the destructive interference among components observed by \cite{nogueira_jfm_2021}. The present results show that such destructive interference occurs among different triadic interactions. Masking all six modes almost entirely eliminates the response as seen in figure \ref{fig:velrecmask}.

\subsection{Energy transfer via triadic interactions} \label{subsec:egy}
The interaction map shown in figure \ref{fig:espmapallfreq}-b can also be interpreted in terms of energy exchange between different modes via nonlinear interactions. \cite{symon_jfm_2021} investigated, by employing the spectral form of the transport equation of turbulent kinetic energy (TKE), the overall relation between production, dissipation and the transfer of energy for individual wavenumber pairs in parallel, stationary turbulent flows. The spectral TKE equation is given, using indicial notation for the last two terms for convenience, as
\begin{align}\label{eq:stke}
\overline{\frac{\partial\hat{E}}{\partial t}}=-\left\langle\frac{\partial\overline{u}}{\partial y}\hat{u},\hat{v}\right\rangle-\frac{1}{Re}\left\langle{\partial {{}\hat{u}_m}}{\partial x_n},\frac{\partial {{}\hat{u}_m}}{\partial x_n}\right\rangle-\left\langle{{}\hat{u}_m},{{}\hat{f}_m}\right\rangle,
\end{align}
where the hat in this equation denotes, by abuse of notation, Fourier transformed quantities in the stremwise and spanwise directions, $\hat{E}$ is the spectral TKE of a given wavenumber pair, the superscript * denotes complex conjugate, $m$ and $n$ denote that the vector indices, $\hat{f}_m$ is the $m^{\textrm{th}}$ component of the forcing vector (see \cite{symon_jfm_2021} for derivation of \eqref{eq:stke}). Here, we assume that the Couette flow is stationary in the time interval we investigate, which renders 
\begin{align} \label{eq:meanstke}
\overline{\frac{\partial \hat{E}}{\partial t}}=0.
\end{align}
The three terms on the right-hand side of \eqref{eq:stke} correspond to the production, dissipation and nonlinear transfer of the turbulent kinetic energy, respectively, which, thanks to \eqref{eq:meanstke}, sum up to zero for a given wavenumber pair. One can write \eqref{eq:stke} in the frequency domain by Fourier transforming in the time domain each term on the right-hand side and Welch averaging \citep{jin_prf_2021}, which still satisfies the energy balance for any wavenumber-frequency triplet $\bm{k}$ as,
\begin{align}\label{eq:stke2}
0=-\left\langle\frac{\partial\overline{u}}{\partial y}\hat{u}_{\bm{k}},\hat{v}_{\bm{k}}\right\rangle-\frac{1}{Re}\left\langle\frac{\partial {{}\hat{u}_m}_{\bm{k}}}{\partial x_n},\frac{\partial {{}\hat{u}_m}_{\bm{k}}}{\partial x_n}\right\rangle-\left\langle{{}\hat{u}_m}_{\bm{k}},{{}\hat{f}_m}_{\bm{k}}\right\rangle.
\end{align}

\begin{figure}
  \centerline{\resizebox{\textwidth}{!}{\includegraphics{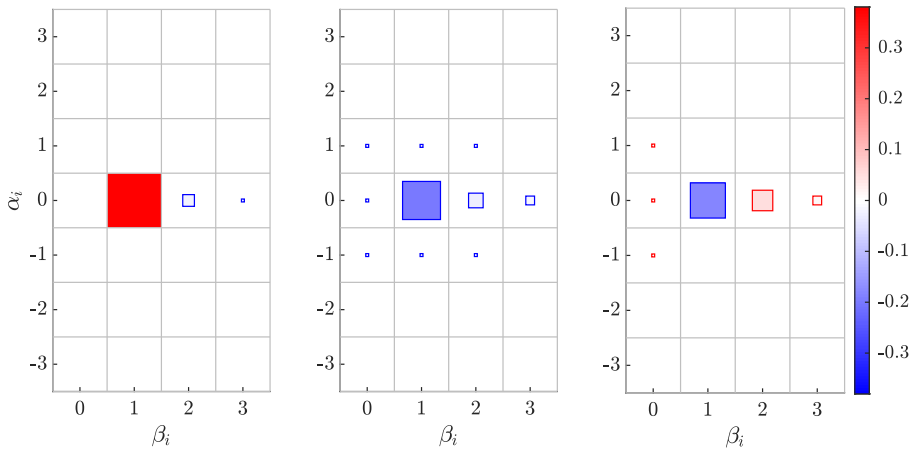}}}
  \caption{Production (left), dissipation (center), and nonlinear transfer (right) of the spectral turbulent kinetic energy for different wavenumber pairs at $\omega=0$. Both the size and the color intensity of the markers indicate amplitude.}
\label{fig:stke}
\end{figure}

The contributions of production, dissipation and nonlinear transfer to the energy balance for different wavenumber pairs at zero frequency are illustrated in figure \ref{fig:stke}. We see that the roll-streak mode (0,1) draws the most energy from the mean flow to produce TKE, and is the only mode to transfer this energy to other modes via nonlinear transfer. All the modes are seen to lose energy via dissipation as expected. Note that the energy balance map is symmetric in the $\beta_i$ axis, which is not shown for better readability of the plot.

\begin{figure}
  \centerline{\resizebox{\textwidth}{!}{\includegraphics{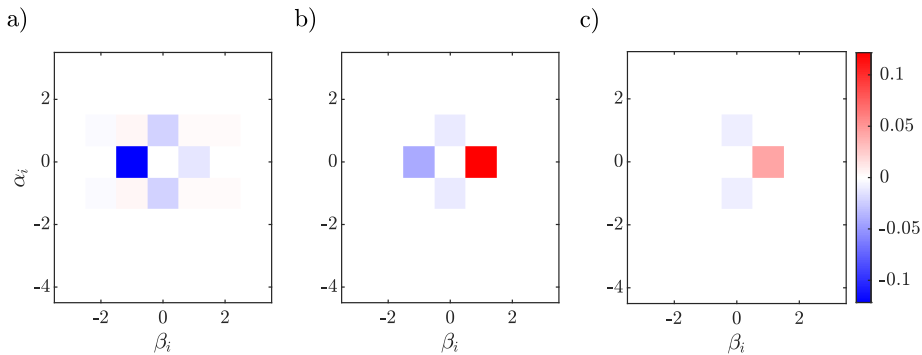}}}
  \caption{Nonlinear energy transfer to the modes (a) $\bm{k}=(0,1,0)$, (b) (0,2,0) and (c) (0,3,0) from different triadic interactions. Transfer from different frequencies for a given wavenumber pair is integrated.}
\label{fig:nlt}
\end{figure}

\cite{cho_jfm_2018} computed the nonlinear energy transfer via each triadic interaction by expanding the convolution in the third term on the right-hand side of \eqref{eq:stke} as in \eqref{eq:conv}. Here, we do a similar analysis for the nonlinear transfer term in the frequency-domain energy balance equation given in \eqref{eq:stke2}. The resulting energy transfer map for (0,1,0) mode is shown in figure \ref{fig:nlt}-a, where blue and red colors indicate losing and gaining energy, respectively, via the corresponding interaction. We see that roll-streak mode (0,1,0) mostly transfer energy via the interactions $(0,\{-1,1\},\omega)+(0,\{2,0\},-\omega)$ and $(\pm 1,0,\omega)+(\mp 1,1,-\omega)$ with $(0,-1,\omega)+(0,2,-\omega)$ being the dominant one. It is also seen to gain energy via a number of interactions but the incoming energy rate is negligible compared to outgoing rate, and hence, is not discussed here. For the triadic interaction $(0,-1,\omega)+(0,2,-\omega)$, we calculated that the transfer from the interaction $(0,-1,0)+(0,2,0)$ is -0.06, which constitutes half of the transfer from all the frequencies. Since the flow is symmetric in the spanwise direction, the modes $(0,1,0)$ and $(0,-1,0)$ are complex conjugates of each other. Thus, we can say that the mode (0,1,0) is transferring energy to roll-streak harmonic mode (0,2,0) via a triadic interaction involving its conjugate mode, making the $(\alpha_k,\beta_k)=(0,2)$ mode the second-most energetic with a peak at zero frequency (see figure \ref{fig:resppsd}). This sort of transfer can be associated to the energy cascade in turbulence from large to small scales observed in high-$Re$ flows. 

Combining the results obtained inspecting the energy transfer and the findings of the previous subsection indicate that the mode (0,1,0) gains almost all of its energy from the mean flow via the lift-up mechanism and transfers some of this energy via nonlinear transfer to streamwise-constant modes as the onset of energy cascade. This nonlinear transfer appears as a streak with opposite phase (see bottom left plot in figure \ref{fig:velrecmask}). This will be further discussed below. The response seen in (0,1,0) mode is a result of the destructive relation between the lift-up mechanism, excited by the wave modes, and the nonlinear transfer associated to roll-streak modes. 


We have seen that (0,2,0) mode receives its energy via a nonlinear transfer mechanism according to the energy budget plot given in figure \ref{fig:stke}. The nonlinear transfer via each triadic interaction contributing to (0,2,0) mode is shown in figure \ref{fig:nlt}-b. The map reveals that (0,2,0) mode receives energy via the interactions $(0,1,\omega)+(0,1,-\omega)$ at a rate similar to the energy transfer in (0,1,0) mode via the interactions $(0,-1,\omega)+(0,2,-\omega)$. Some of the energy that (0,2,0) mode receives is transferred to the next harmonic via the interaction $(0,-1,\omega)+(0,3,\omega)$. Looking at the nonlinear transfer map for (0,3,0) mode in figure \ref{fig:nlt}-c, we see that energy is transferred via the interaction $(0,1,0)+(0,2,0)$, once again, at a rate similar to the transfer from (0,2,0) shown in figure \ref{fig:nlt}-b. One can trace the energy cascade for the negative $\beta$ modes in the same way, which yields the same transfer maps mirrored in the $\alpha_i$ and $\beta_i$ axes. This suggests that the modes $(\alpha,\beta)=(0,\pm1)$, once extracting energy from the mean and transferring it to harmonics $(0,\pm2)$, plays a role to transfer energy via the triadic interactions associated to higher $\beta$, i.e., they provide a medium for the nonlinear transfer to higher $\beta$ modes without losing noticeable energy.

\begin{figure}
  \centerline{\resizebox{\textwidth}{!}{\includegraphics{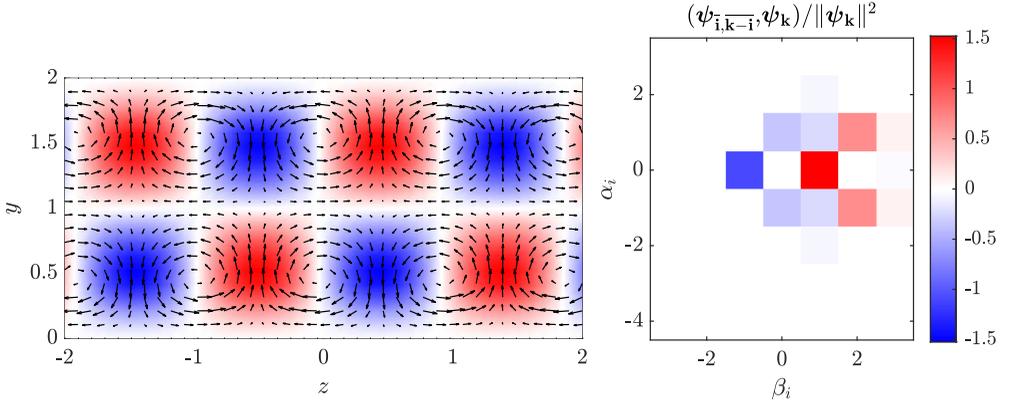}}}
  \caption{The same map with figure \ref{fig:espmapallfreq}-b obtained for the mode $(\alpha_k,\beta_k,\omega_k)=(0,2,0)$ (right) and the corresponding response in the $y$-$z$ plane (left).}
\label{fig:espmapaf002}
\end{figure} 

\begin{figure}
  \centerline{\resizebox{\textwidth}{!}{\includegraphics{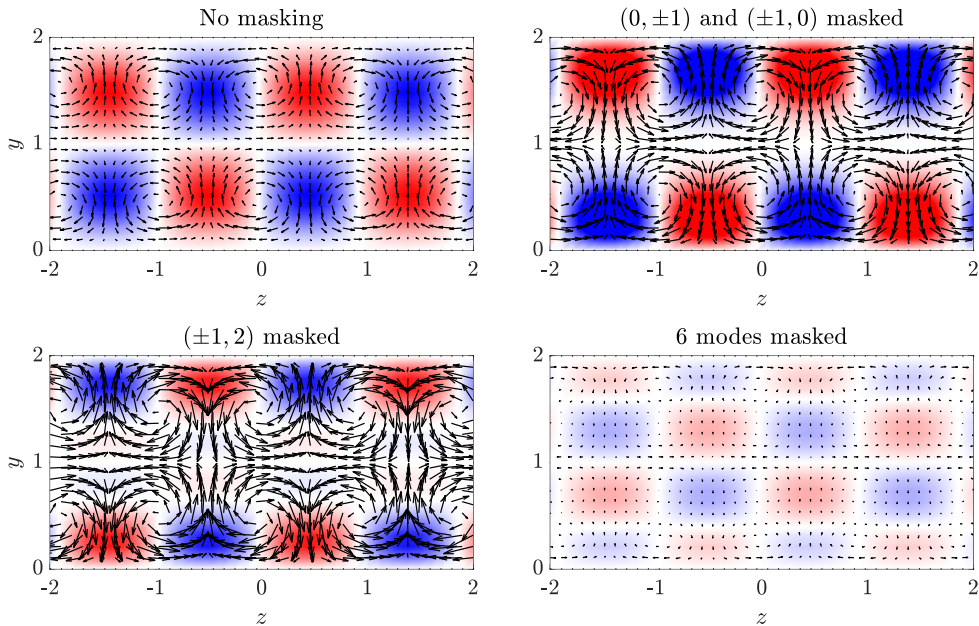}}}
  \caption{Velocity field corresponding to the optimal SPOD mode at $(\alpha_k,\beta_k,\omega_k)=(0,2,0)$ in the $y$-$z$ plane. Top-left: the entire response; top-right: response obtained by masking the interactions $(\alpha_i,\beta_i)=(0,\pm1)+$complementary and $(\pm1,0)+$complementary; bottom-left: response obtained by masking the interactions $(\pm1,2)+$complementary; bottom-right: the response obtained by masking the six interactions.}
\label{fig:velrecmask2}
\end{figure} 

We now investigate whether a destructive relation takes place between the nonlinear transfer and the lift-up mechanism for roll-streak harmonic mode $(0,2,0)$ as in roll-streak mode $(0,1,0)$. In figure \ref{fig:espmapaf002}, we show the same interaction map given in figure \ref{fig:espmapallfreq}-b for the mode $(\alpha_k,\beta_k,\omega_k)=(0,2,0)$ together with the reconstruction of the mode in the $y$-$z$ plane. Similar to the $(0,1,0)$ mode, we see streamwise vortices and streaks in $(0,2,0)$ with a halved period in the spanwise direction. The rolls and the streaks are in opposite phase compared to the $(0,1,0)$, which implies that they are not directly associated with the lift-up mechanism. The interaction map indicates strong contribution to response generation from the interactions. Besides these interactions, we see some positive contributions from the interactions $(\pm1,2)+(\mp1,0)$. To see the effect of these interactions on the response, we mask these interactions and observe the change in the response in figure \ref{fig:velrecmask2}. Masking the interactions associated to nonlinear transfer, we obtain a response field reminiscent of the lift-up mechanism. This partial response is due to the interactions $(\pm1,2)+(\mp1,0)$. Masking these interactions, on the other hand, yields a response field with inverted streaks and vortices. Similar to the case of $(0,1,0)$ mode, there exists a destructive interference between the nonlinear energy transfer and the lift-up mechanism. Masking both groups of interactions causes the response to be almost zero, indicating that these six interactions are the active ones for response generation.

\section{Source-sink decomposition and modelling sinks using eddy viscosity} \label{sec:eddy}
\subsection{Couette flow at $Re=400$}
{Our analysis so far reveals that different nonlinear interactions can have constructive or destructive effects on a given flow structure as seen in the interaction maps given in figures \ref{fig:espmapallfreq}-b and \ref{fig:espmapaf002}. Guided by these results, we now investigate how to combine eddy viscosity with the resolvent framework when forcing data is involved.} 

For a given observed mode, we separate the triadic interactions into sources and sinks, depending on whether they have a constructive or destructive effect on the mode, respectively. We then model the sinks using eddy viscosity when constructing the resolvent operator and use the sources to drive this modified resolvent operator. The equation describing this procedure reads
\begin{align} \label{eq:respodsrc}
{\bm{\psi}}_{\bm{k}}\approx\tilde{\mathsfbi{R}}_{\bm{k}}{\bm{\chi}}_{\bm{k}}^+,
\end{align}
where $\tilde{\mathsfbi{R}}_{\bm{k}}$ denotes the resolvent operator constructed using eddy viscosity and ${\bm{\chi}}_{\bm{k}}^+$ denotes the RESPOD forcing mode constructed using only the interactions with a positive impact on ${\bm{\psi}}_{\bm{k}}$, i.e., the interactions depicted with a red square in figure \ref{fig:espmapallfreq}-b. Note that this equation is not exact, unlike \eqref{eq:respod}. This suggests that a successful prediction of the flow field using \eqref{eq:respodsrc} indicates that the source/sink decomposition proposed here points to a physical mechanism about how nonlinear interactions saturate the turbulent fluctuations. 

\begin{figure}
  \centerline{\resizebox{\textwidth}{!}{\includegraphics{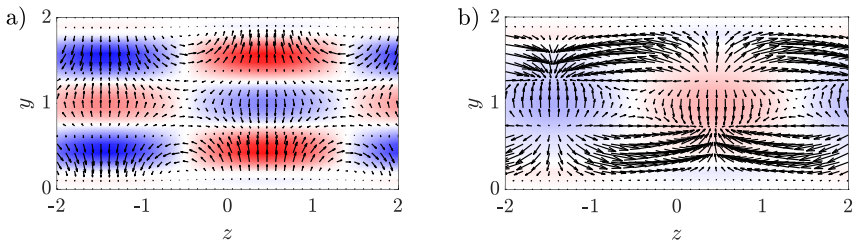}}}
  \caption{The RESPOD forcing mode, ${\bm{\chi}}_{\bm{k}}$, for $\bm{k}=(\alpha_k,\beta_k,\omega_k)=(0,1,0)$ decomposed into (a) sink, ${\bm{\chi}}_{\bm{k}}^-$, and (b) source, ${\bm{\chi}}_{\bm{k}}^+$.}
\label{fig:frcrecsrc}
\end{figure}

In figure \ref{fig:frcrecsrc}, we show the source-sink decomposition of the RESPOD forcing mode for $\bm{k}=(0,1,0)$ (see figure \ref{fig:frcrespmode}-b). We see that the streamwise vortex seen in figure \ref{fig:frcrespmode}-a is generated by the $y$- and $z$-components of the source, $\bm{\chi}_{\bm{k}}^+$. The sink, on the other hand, contains mostly a streamwise component which has its peak towards the wall and is in opposite phase with the streak in \ref{fig:frcrespmode}-a. This suggests that the source is associated with the lift-up mechanism, creating a roll-streak structure, whose energy is dissipated to some extent by the $x$-component of the sink.

We investigate the use of eddy viscosity to model the effect of the sink. The eddy viscosity, $\nu_t$, is obtained using the Cess model \citep{cess_1958} given as
\begin{align}
\frac{\nu_t}{\nu} = \frac{1}{2}\left(1+\frac{\kappa^2Re_\tau^2}{9}(1-y^2)^2(1+2y^2)^2(1-e^{-Re_\tau(1-|y|)/A})^2\right)^{1/2}-\frac{1}{2},
\end{align}
where $\nu$ denotes the molecular viscosity, the von K\'arman constant $\kappa=0.426$ and the constant $A=25.4$ as in \cite{alamo_jfm_2006}. To construct the modified resolvent operator, the molecular viscosity in $\mathsfbi{R}_{\bm{k}}$ is replaced by $\nu_t$. Note that $\nu_t$ is dependent on $y$ which should be accounted for when differentiating the viscosity terms in $\mathsfbi{R}_{\bm{k}}$ \citep[see ][]{hwang_jfm_2010}. 

\begin{figure}
  \centerline{\resizebox{\textwidth}{!}{\includegraphics{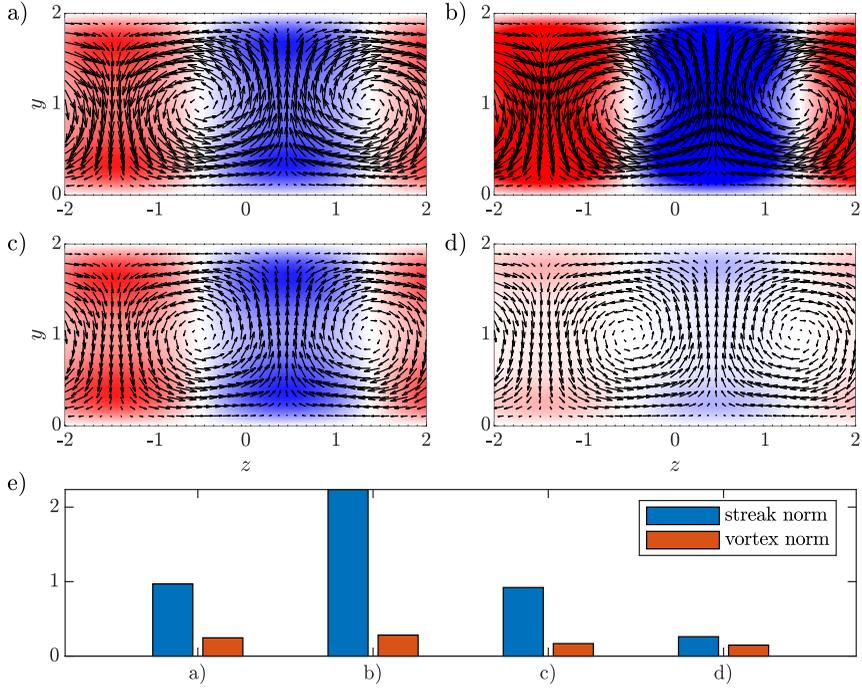}}}
  \caption{The optimal SPOD mode for $(\alpha_k,\beta_k,\omega_k)=(0,2,0)$ (a) compared against the response prediction obtained using (b) ${\mathsfbi{R}}_{\bm{k}}{\bm{\chi}}_{\bm{k}}^+$, (c) $\tilde{\mathsfbi{R}}_{\bm{k}}{\bm{\chi}}_{\bm{k}}^+$ and (d) $\tilde{\mathsfbi{R}}_{\bm{k}}{\bm{\chi}}_{\bm{k}}$; (e) norms in each case for $u$ and $[v\;w]^\top$ components, which correspond to the streaks and streamwise vortices, respectively. }
\label{fig:velrecsrc}
\end{figure}

In figure \ref{fig:velrecsrc}, the analysis is tested for $(0,1,0)$ mode, which is the dominant wavenumber-frequency triplet. 
 The flow field obtained using \eqref{eq:respodsrc} is compared against the optimal SPOD mode of the response ${\bm{\psi}}_{\bm{k}}$ as well as the response fields generated by ${\mathsfbi{R}}_{\bm{k}}{\bm{\chi}}_{\bm{k}}^+$ and $\tilde{\mathsfbi{R}}_{\bm{k}}{\bm{\chi}}_{\bm{k}}$. As expected, masking the sinks from the forcing causes an overprediction of the response in the case of ${\mathsfbi{R}}_{\bm{k}}{\bm{\chi}}_{\bm{k}}^+$. On the other hand, using the RESPOD forcing mode without any decomposition to drive the modified resolvent operator ($\tilde{\mathsfbi{R}}_{\bm{k}}{\bm{\chi}}_{\bm{k}}$) leads to underprediction of the response due to the enhanced damping in $\tilde{\mathsfbi{R}}_{\bm{k}}$. {To quantify the accuracy of each prediction, the norms for the streak ($u$) and the streamwise vorticity ($[v\;w]^\top$) components of the predicted response are also compared against those for the optimal SPOD mode. We see that removing the sinks and using eddy viscosity with full forcing mostly affect the streaks, doubling and halving its norm, respectively.}  When \eqref{eq:respodsrc} is used instead, these two effects, i.e., the amplification due to the use of sources only and the enhanced damping due to the use of eddy viscosity, almost perfectly cancel each other, leading to an accurate response prediction. 

\subsection{Couette flow at $Re=1000$}
\begin{figure}
  \centerline{\resizebox{\textwidth}{!}{\includegraphics{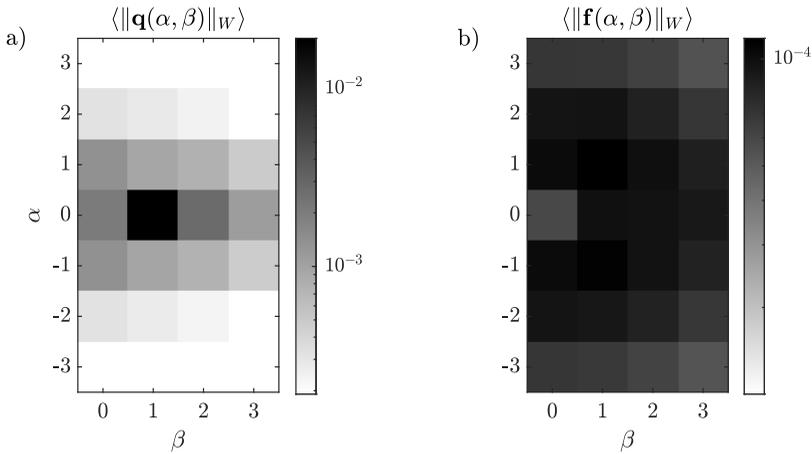}}}
  \caption {Time average of the energy of flow structures (a) and the associated forcing (b) at different wavenumber pairs at $Re=1000$. The color-scale ranges two and one order of magnitude for plots (a) and (b), respectively.}
\label{fig:resprmshigh}
\end{figure}

We now consider the Couette flow at $Re=1000$. The energy distribution per wavenumber pair and the associated forcing is shown in figure \ref{fig:resprmshigh}. Once again, the dominant mode is seen to be $(\alpha,\beta)=(0,1)$ while the energy of the modes with higher wavenumbers are seen to be larger compared to the $Re=400$ case, which is expected due to the enhanced turbulence activity and energy transfer to smaller scales due to the increase of $Re$. The energy of the forcing is more evenly distributed towards smaller scales with its peak at the $(1,\pm1)$ mode pair. 

\begin{figure}
  \centerline{\resizebox{\textwidth}{!}{\includegraphics{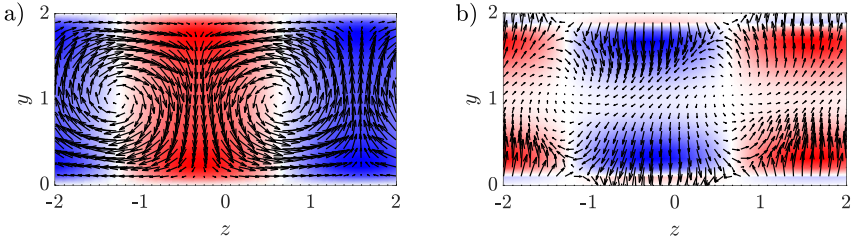}}}
  \caption{The optimal SPOD mode $\bm{\psi}_{\bm{k}}$ (a) and the associated forcing $\bm{\chi}_{\bm{k}}$ (b) for the mode $\bm{k}=(\alpha,\beta,\omega)=(0,1,0)$ at $Re=1000$.}
\label{fig:frcrespmodehigh}
\end{figure} 

\begin{figure}
  \centerline{\resizebox{\textwidth}{!}{\includegraphics{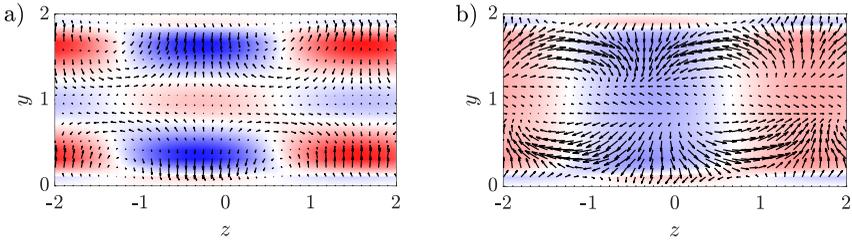}}}
  \caption{The RESPOD forcing mode for $\bm{k}=(\alpha_k,\beta_k,\omega_k)=(0,1,0)$ at $Re=1000$ decomposed into (a) sink, ${\bm{\chi}}_{\bm{k}}^-$, and (b) source, ${\bm{\chi}}_{\bm{k}}^+$.}
\label{fig:frcrecsrchigh}
\end{figure} 

We test the source-sink decomposition approach at $(\alpha,\beta,\omega)=(0,1,0)$, which is the dominant mode similar to the $Re=400$ case. The optimal SPOD mode, $\bm{\psi}_{\bm{k}}$, and the associated RESPOD forcing mode, $\bm{\chi}_{\bm{k}}$, are given in figure \ref{fig:frcrespmodehigh}. Once again, we see a roll-streak mode in $\bm{\psi}_{\bm{k}}$ reminiscent of the lift-up mechanism. The associated forcing, whose $x$-component is in opposite phase of the streak and the $y$- and $z$- components are mostly aligned with the streamwise vortex, is decomposed into a source and a sink as shown in figure \ref{fig:frcrecsrchigh}. Similar to the $Re=400$ case, we observe that the $y$- and $z$-components of the source generates the streamwise vortex, while the $x$-component of the sink is responsible for reducing the streak energy. 

\begin{figure}
  \centerline{\resizebox{\textwidth}{!}{\includegraphics{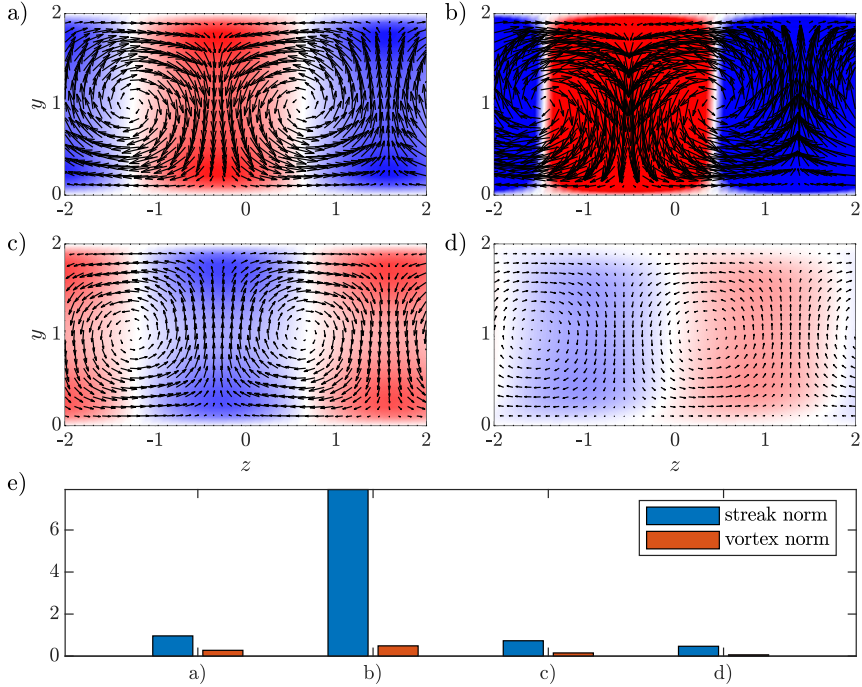}}}
  \caption{Same as figure \ref{fig:velrecsrc}; demonstrating the $Re=1000$ case.}
\label{fig:velrecsrchigh}
\end{figure} 

We again test modelling the effect of the sink via eddy viscosity. For the $Re=1000$ flow, the sensitivity to any imbalance between the sources and the sinks is significantly higher due to the stronger linear amplification mechanisms associated with higher $Re$. {This is clearly observed in figure \ref{fig:velrecsrchigh}-b and \ref{fig:velrecsrchigh}-e where omitting the sinks when driving the original resolvent operator, ${\mathsfbi{R}}_{\bm{k}}$, yields a strong overprediction of the response ($\sim7$ folds the SPOD norm), which is much more evident compared to the $Re=400$ case ($\sim2$ folds). Driving the modified resolvent operator, $\tilde{\mathsfbi{R}}_{\bm{k}}$, with sources and sinks together causes an underprediction of the response due to the enhanced viscous damping as seen in figure \ref{fig:velrecsrchigh}-d, halving the norm. Note that given the difference in the over- and underprediction ratios in figure \ref{fig:velrecsrchigh}-e for cases shown in -b and -d, one can conclude that inclusion of eddy viscosity improves the robustness of the resolvent-based prediction by reducing the operator sensitivity to imbalance between the sources and sinks. Once again, we see that these two effects cancel each other leading to the closest prediction of the response as seen in figure \ref{fig:velrecsrchigh}-c with a 30\% underprediction of the norm.}

\begin{figure}
  \centerline{\resizebox{\textwidth}{!}{\includegraphics{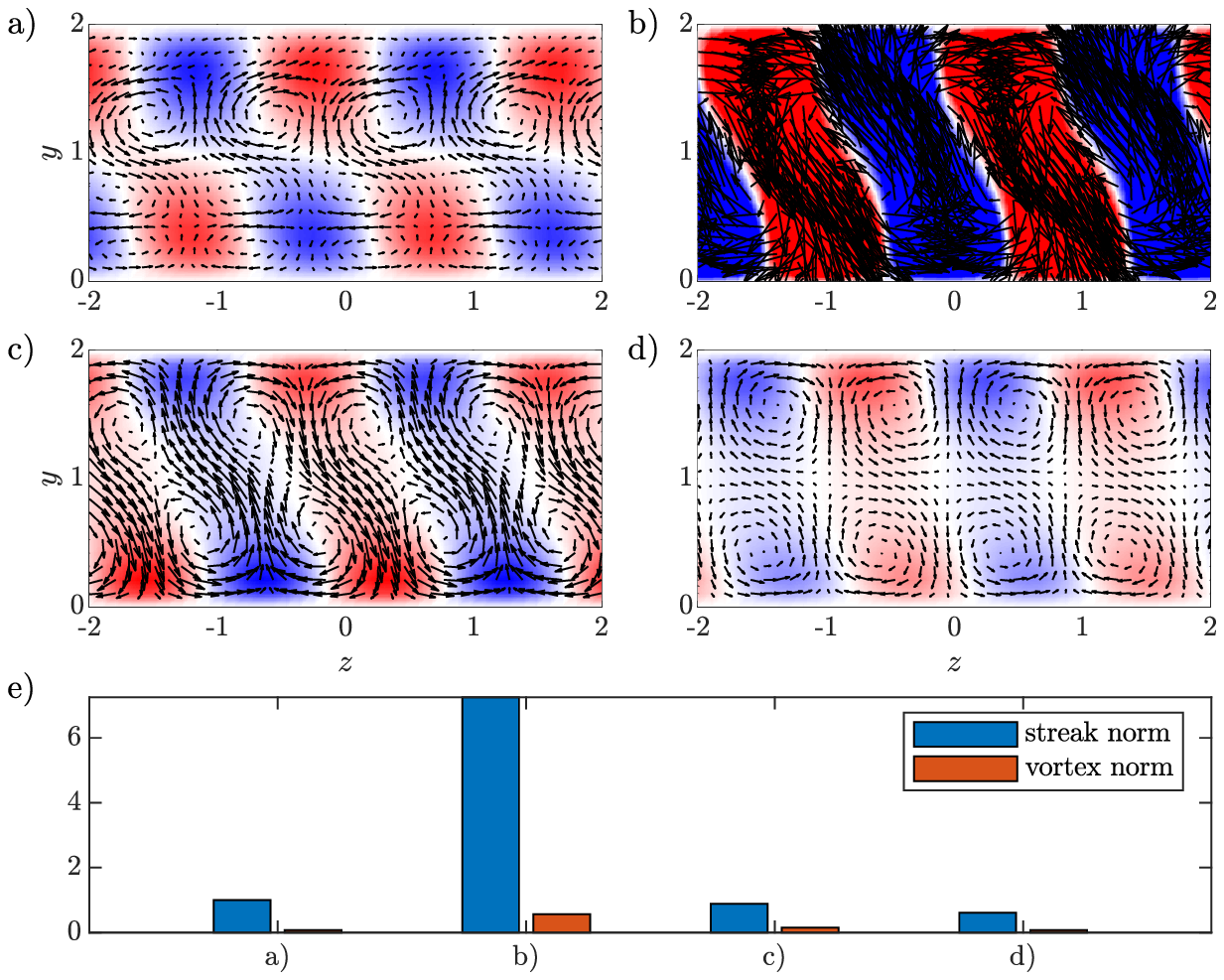}}}
  \caption{Same plot as figure \ref{fig:velrecsrchigh}; demonstrating the mode $(\alpha,\beta,\omega)=(0,2,0)$.}
\label{fig:velrecsrchigh2}
\end{figure}

\begin{figure}
\centerline{\resizebox{\textwidth}{!}{\includegraphics{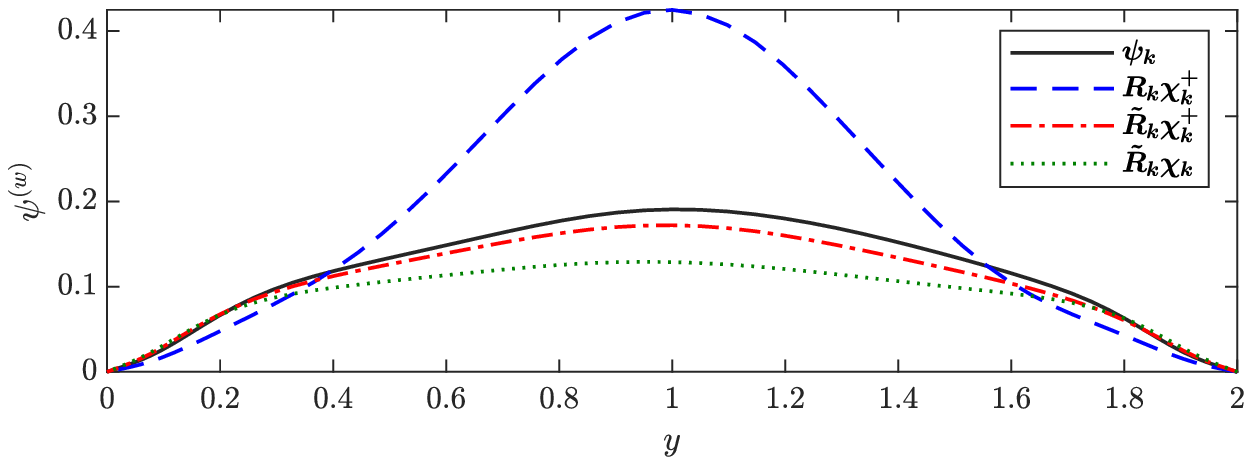}}}
\caption{The spanwise component of the optimal SPOD mode (solid, black) compared against ${\mathsfbi{R}}_{\bm{k}}{\bm{\chi}}_{\bm{k}}^+$ (red, dashed), $\tilde{\mathsfbi{R}}_{\bm{k}}{\bm{\chi}}_{\bm{k}}^+$ (blue, dash-dotted) and $\tilde{\mathsfbi{R}}_{\bm{k}}{\bm{\chi}}_{\bm{k}}$ (green, dotted) at $\bm{k}=(\alpha,\beta,\omega)=(1,0,0)$.}
\label{fig:velrecsrchigh3}
\end{figure} 
Finally, we show the results of source-sink decomposition in figures \ref{fig:velrecsrchigh2} and \ref{fig:velrecsrchigh3}for $(0,2,0)$ and $(1,0,0)$ modes, which are the second and third most dominant modes, respectively. Only the spanwise component is shown for the $(1,0,0)$ mode since the other two components tend to zero. Similar to the results for $(0,1,0)$ mode, leaving out the sinks or adding eddy viscosity results in over- and underprediction of the response, respectively, while the effect of these two modifications are mostly neutralised when applied at the same time.
\subsection{Discussion on the usefulness of source-sink decomposition}

This analysis elucidates how eddy viscosity modelling can be integrated into resolvent analysis in the presence of forcing data to make quantitative predictions. When forcing data is present, the use of eddy viscosity to predict the flow may seem redundant. However, it may actually be an interesting approach due to several reasons. 

Generating forcing data accurate enough to satisfy \eqref{eq:resmain} with minimal errors for high-$Re$ turbulent flows is a difficult task, as we discussed in detail in an earlier study \citep{karban_tcfd_2022}. Modelling some of the nonlinear interactions with eddy viscosity can significantly enhance the robustness of the use of resolvent framework with forcing data. 

Besides, this framework is also advantageous for modelling purposes. In \cite{karban_jfm_2022}, we showed that the self-similarity of turbulent structures, which is based on Townsend's attached eddy hypothesis \citep{townsend_mp_1951,townsend_jfm_1976}, extends to the forcing terms, implying self-similar turbulence generation mechanisms. This suggests that it may be possible to devise self-similar forcing models for high-$Re$ flows. However, using such models to drive the resolvent operator to make quantitative response predictions is a hard task due to the sensitivity of the resolvent operator to any inaccuracy in the forcing model. The framework we present here helps overcome this issue in two ways: (i) one only needs to model the sources as the sinks are modelled via eddy viscosity, and (ii) the enhanced damping in the modified resolvent operator allows more tolerance to errors in the forcing model. Therefore, we believe that this framework opens significant room for future progress in quantitative flow modelling using the resolvent framework.

\section{Conclusions} \label{sec:conc}
We have discussed a method to investigate the triadic interactions that underpin the generation of flow structures associated with a given observable. The method is based on the resolvent-based extended spectral proper orthogonal decomposition (RESPOD), used in \cite{karban_jfm_2022} to identify self-similar structures in a turbulent channel flow. A minimal Couette flow is here chosen as the test case, where the triadic interactions associated with optimal SPOD modes are investigated. 

{We identify the forcing modes correlated to the SPOD modes of an observable via RESPOD. These forcing modes generates the associated SPOD modes when applied to the resolvent operator. We show in this study that using RESPOD, it is also possible to identify individual triadic interactions that are correlated to the observable. Summation of the correlated triadic interactions is by definition equal to the RESPOD forcing mode. This procedure allows identifying interactions that dominate generating the observable.}

Our analysis reveals that the most energetic mode, $(\alpha,\beta)=(0,1)$, at its peak-energy frequency, $\omega=0$, is mainly driven by six triadic interactions: four interactions involving modes periodic over $L_x$ in the streamwise direction, that generate small-in-amplitude but efficient forcing, and two interactions involving streamwise-constant modes that, although being less efficient, generate forcing structures with large amplitudes. The streamwise-periodic interactions generate a combined streak-streamwise vortex structure via the lift-up mechanism, while the streamwise-constant interactions counteract the streak generation by generating a streamwise forcing component in phase opposition to the lift-up mechanism. This explains in physical terms the destructive interference of forcing observed by \cite{nogueira_jfm_2021}: forcing is composed of different triadic interactions with opposing effects in exciting streamwise vortices and streaks.

Our framework also allows us to investigate energy transfer between different modes via triadic interactions. We observe that the triadic interactions involving the $(0,1)$ mode provide a constructive contribution to all the modes investigated. This is an expected result since it is the only mode with a negative nonlinear transfer rate of turbulent kinetic energy, as shown by the energy balance analysis we conducted following \cite{symon_jfm_2021}. Investigating the nonlinear transfer for different modes via a range of triadic interactions, we observe the energy cascade mechanism transferring energy from $(0,1)$ to $(0,2)$ and then from $(0,2)$ to $(0,3)$. Comparing the analyses based on the resolvent framework and energy budget revealed that the triadic interactions associated to the lift-up mechanism and the nonlinear transfer are in destructive interference for the modes $(0,1,0)$ and $(0,2,0)$.  

{To demonstrate how the resolvent-based framework we developed here is associated with eddy viscosity modelling, we introduced a combined modelling strategy: we group the nonlinear interactions as sources and sinks, then, model the sinks using eddy viscosity when constructing the resolvent operator, and drive this modified resolvent operator with the sources only. Our tests in the Couette flow at $Re=400$ and 1000 revealed that such a decomposition, despite being ad-hoc, yields an accurate response prediction. This elucidates how to incorporate eddy viscosity modelling into the resolvent framework in the presence of forcing data.}

{Finally, we discussed that such a modelling framework could be beneficial for using resolvent analysis to make quantitative predictions for high-$Re$ turbulent flows. The modified resolvent operator offers enhanced damping thanks to the use of eddy viscosity, which reduces its sensitivity to errors in the forcing. One can attempt modelling the sources only and use this modelled forcing to drive the modified resolvent operator, which will be to subject of a future study. We only showed results for modes at zero frequency limit, since these modes dominate the Couette flow. To further explore the potential of the method introduced, analysis of different flows at higher $Re$ will be conducted.}

\backsection[Funding]{This work has received funding from the Clean Sky 2 Joint Undertaking under the European Union’s Horizon 2020 research and innovation programme under grant agreement No 785303. U.K. has received funding from TUBITAK 2236 Co-funded Brain Circulation Scheme 2 (Project No: 121C061).}

\backsection[Declaration of interests]{The authors report no conflict of interest.}

\bibliographystyle{jfm}
\bibliography{biblio}

\end{document}